\documentstyle[aps,epsfig,float,twocolumn]{revtex}
\begin{document}

\title{Strong Correlation to Weak Correlation Phase Transition\\ 
in Bilayer Quantum Hall Systems}
\author{John Schliemann, S.~M. Girvin, and A.~H. MacDonald}
\address{Department of Physics, Indiana University, Bloomington, IN
47405-7105}
\date{\today}
\maketitle
\begin{abstract}
At small layer separations, the ground state of a $\nu=1$ bilayer
quantum Hall system exhibits spontaneous interlayer phase coherence.
The evolution of this state with 
increasing layer separation $d$ has been a matter of controversy.   
We report on small system exact diagonalization
calculations which suggest that a single phase transition,
likely of first order, separates incompressible  
states with strong interlayer correlations from compressible
states with weak interlayer correlations.  We find a dependence of the phase 
boundary on $d$ and interlayer tunneling amplitude that is in very good 
agreement with recent experiments.
\end{abstract}

\vspace{0.5cm}

The ground state of a two-dimensional monolayer electron system at Landau 
level filling factor $\nu=1$ is a single Slater determinant described exactly 
by Hartree-Fock theory and is   %SMG
a strong ferromagnet with a 
large gap $E_g$ for charged excitations \cite{DaPi:97,halperin83}.
This elementary property has rich and interesting consequences for
the physics of bilayer quantum Hall systems at the same total $\nu$,
consequences 
that are readily appreciated when a pseudospin language \cite{MPB:90,DaPi:97}
is used to describe the layer degree of freedom. 
When the layer separation $d$ goes to zero, interactions between layers
are identical to interactions within  
layers.  The pseudospin bilayer Hamiltonian is then identical to the
single layer Hamiltonian with spin and its ground
state has pseudospin order and a finite charge gap.
For infinite layer separation,
on the other hand, the bilayer system reduces to two disordered, 
compressible, uncorrelated $\nu=1/2$ systems. 
This Letter concerns the evolution of bilayer quantum Hall systems 
between these two extremes.

For small layer separations the difference bet\-ween interlayer and 
intralayer interactions breaks 
the pseudospin-invariance of the Hamiltonian, resulting in an
incompressible easy-plane pseudospin ferromagnet.  
In physical terms the pseudospin order represents spontaneous
phase coherence between the electron layers.
Several scenarios have been proposed for the evolution of the ground state 
as the layer separation increases further. In Hartree-Fock theory 
\cite{HF:DBM}, spontaneous interlayer coherence is lost if the layer 
separation exceeds a critical value, and the ground state 
at large separations consists of weakly correlated 
Wigner crystals. While possibly instructive, 
this picture is known to be incorrect at large $d$ since
half-filled Landau levels do hot have crystalline ground states.
Working in the other direction, Bonesteel {\it et al.} started 
\cite{bonesteel} 
from the composite fermion theory of isolated compressible 
$\nu=1/2$ layers, and concluded that coupling would lead to pairing between 
composite fermions in opposite layers and also, implicitly, to a charge gap.
Since the pseudospin ferromagnet possesses particle-hole rather 
than particle-particle pairing, however, this picture 
still implies that at least one phase transition occurs as a function 
of layer separation.  In a numerical diagonalisation study
He {\it et al.} \cite{HDX:93} predicted, on the basis of 
the system parameter dependence of 
overlaps between exact groundstates and two different variational
wavefunctions, the existence of two distinct incompressible states 
separated by a region of compressible states.

Experiments, on the other hand, have tended to be 
consistent \cite{MEBPW:94} with the proposal \cite{MPB:90} that a single-phase 
transition from an incompressible to a compressible states occurs
with increasing layer separation at any value of the 
interlayer tunneling amplitdue.
Very recently, in an intriguing new experiment by Spielman {\it et al.} 
\cite{SEPW:00}.
the tunneling conductance across the layers was studied in a 
sample with extremely small tunneling amplitude. 
When the ratio of layer separation and magnetic length was lowered (at fixed
filling factor) below a 
critical value, the conductance showed
a very pronounced peak around zero bias voltage between the layers,
that provides direct evidence \cite{recentcondmats} for spontaneous 
interlayer phase coherence.  
This is because in the coherent state, the layer 
index of each electron is uncertain and only in this case can tunneling
leave the system in or near its ground state so that there is no orthogonality
catastrophe and tunneling can occur at zero voltage.

Since the critical layer separation found by Spielman {\it et al.} is close
to the one obtained earlier by Murphy {\it et al.} for the onset of the 
quantum Hall effect \cite{MEBPW:94},
experiment demonstrates that for vanishing tunneling amplitude
the phase transitions at which pseudospin order
and the charge gap are lost are either closely spaced or 
coincident.  

In this Letter we report on small
system exact diagonalization calculations which strongly suggest 
that bilayer quantum Hall systems have a single phase transition,
likely of first order, as a function of $d$. Our critical layer separation  
is in very good quantitative agreement with the value measured %SMG
in Ref.  \cite{SEPW:00}. 
In the light of the experimental results mentioned above, our calculations 
imply that the charge gap disappears and long-range phase coherence 
simultaneously drops sharply to near zero at the phase transition.  
This result is not entirely unexpected since
a simple Landau-Ginzburg analysis indicates that the two order parameters
could not vanish simultaneously without fine-tuning, if the transition were 
continuous.  Also the mean-field theory energy gap is proportional to
the pseudospin order parameter, suggesting that these two orders 
are mutually reinforcing and that a first order transition is therefore likely.
Finally, we note that, experimentally, the charge gap phase
transition is sharp even at finite tunneling between the layers.  Since
tunneling produces a pseudomagnetic field which
couples to the pseudospin order parameter,  this is an unusual
magnetic transition which does {\em not} involve symmetry breaking, a fact
which lends further weight to the suggestion that the transition is first
order. 

We analyse bilayer quantum Hall systems numerically by means of exact
diagonalisations of finite systems unsing the spherical geometry.
We have verified numerically that the ground state and low-lying 
excitations are fully spin-polarized and neglect the spin degree
of freedom in the present discussion.
The Hamiltonian is given by
\begin{equation}
{\cal H}={\cal H}_{{1\rm P}}+{\cal H}_{{\rm Coul}}\quad,
\label{defmod}
\end{equation}
where ${\cal H}_{{\rm Coul}}$ represents the usual Coulomb interaction
within and between layers, and the single-particle Hamiltonian 
${\cal H}_{{1\rm P}}$ is given by
\begin{equation}
{\cal H}_{{1\rm P}}=
-\frac{1}{2}\sum_{m}c^{+}_{\mu,m}
\big[\Delta_{v}\tau^{z}_{\mu,\mu'} 
+\Delta_{t}\tau^{x}_{\mu,\mu'} \big]
c_{\mu',m}\,.
\end{equation}
We concentrate here on the tunneling amplitude 
($\Delta_{t}$) tuned phase transition, although bias 
voltage ($\Delta_{v}$) dependence is also interesting and 
often experimentally more convenient. 
$\mu,\mu'\in\{+,-\}$ run over the layer (or pseudospin) indices and a 
summation convention is implicit; $\vec\tau$ are the pseudospin
Pauli matrices. 
$m\in\{-N_{\phi}/2,\dots,N_{\phi}/2\}$
is the z-projection of the orbital angular momentum of each electron
in the lowest Landau level, 
where $N_{\phi}$ is the number of flux quanta penetrating the sphere.
In the following we
denote the pseudospin operators by
$\vec T=(1/2)\sum_{m}c^{+}_{\mu,m}
\vec\tau_{\mu,\mu'} c_{\mu',m}$.
The interlayer separation $d$ is measured in units of the magnetic length
$l_{B}=\sqrt{{\hbar c}/{eB}}$, and all energies are given in units of the
Coulomb energy scale $e^{2}/\epsilon l_{B}$.
We consider the case of
zero well width to enable comparison with most previous theoretical
investigations\cite{MMYGMZYZ:95,YMBMGMZY:96,Moo:97,JoMa:99}), and 
also systems consisting of two rectangular wells of finite width $w$ 
\cite{note1} whose
ratio to the center-to-center layer separation $d$ is $w/d=0.65$. This value
corresponds to the sample used in Ref. \cite{SEPW:00}.

We consider systems with an even electron number
$N$ which leads to a nondegenerate spatially homogeneous ground state with 
total angular momentum $L=0$. For simplicity, let us first examine the case of
vanishing bias voltage, where both $\langle T^{y}\rangle$ and 
$\langle T^{z}\rangle$ are strictly zero.

\begin{figure}
\begin{center}
\centerline{\includegraphics[width=8cm]{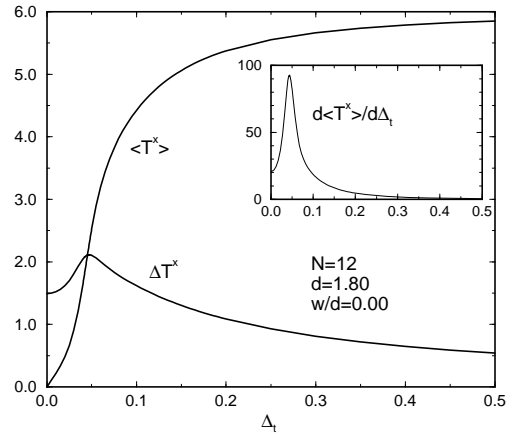}}
\caption{The pseudospin expectation value 
$\langle T^{x}\rangle$ and the fluctuation $\Delta T^{x}$
as a function of the tunneling gap $\Delta_{t}$.
The derivative $d\langle T^{x}\rangle/d\Delta_{t}$ (measured in units of
$1/(e^{2}/\epsilon l_{B})$) is shown in the inset.
\label{fig1}}
\end{center}
\end{figure}

\begin{figure}
\begin{center}
\centerline{\includegraphics[width=8cm]{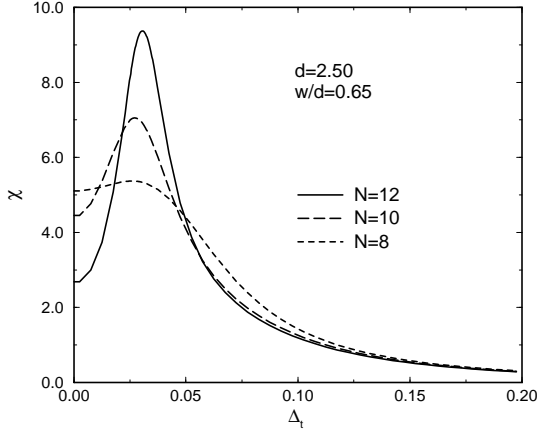}}
\caption{The pseudospin susceptibility $\chi$ for different system sizes
as a function of the tunneling gap. The rapidly growing peak indicates a
quantum phase transition.
\label{suszifig}}
\end{center}
\end{figure}

Figure \ref{fig1} shows 
the interlayer phase coherence as measured by
the expectation value $\langle T^{x}\rangle$
along with the fluctuation $\Delta T^{x}=
\sqrt{\langle{T^{x}}^{2}\rangle-\langle T^{x}\rangle\langle T^{x}\rangle}$ 
as a function of the tunneling gap 
for a system of twelve electrons, a layer separation of $d=1.80$, and zero
well width. At $\Delta_{t}=0$, $\langle T^{x}\rangle$ is necessarily zero in a 
finite system. With increasing tunneling gap, $\langle T^{x}\rangle$ grows
rapidly reaching a inflection point with a very steep tangent. The 
differential pseudospin susceptibility, 
$\chi = (1/N) \langle T^{x}\rangle/d\Delta_{t}$, is
plotted in the inset and shows a very pronounced peak.  
In the immediate vicinity of this peak, the pseudospin
fluctuation $\Delta T^{x}$ has also a pronounced maximum. 
In figure \ref{suszifig} the $\chi$ is plotted for 
different numbers of electrons. 

The rapid growth with increasing 
system size of the peak in this generically intensive quantity 
is strong evidence for a ground state phase transition. Analogous
findings are obtained for the peak in the pseudospin fluctuation.
Thus, the peaks in the susceptibility of the pseudospin and its
fluctuation grow very rapidly with increasing system size  and signal a 
quantum phase transition at the critical value of the tunneling gap. At large 
tunneling the system pseudospin magnetisation
is close to its maximum value, while at small (but also finite) tunneling the
the system is disordered and the pseudospin magnetisation is 
strongly reduced by interactions.

The two peaks described above occur at extremely nearby values of $\Delta_{t}$
at a given layer separation $d$,
and we consider the very tiny differences in their location as a finite-size
effect. To estimate the phase diagram of the system we place the
phase boundary at the maximum of the quantum fluctuations $\Delta T^{x}$. 

\begin{figure}
\begin{center}
\centerline{\includegraphics[width=8cm]{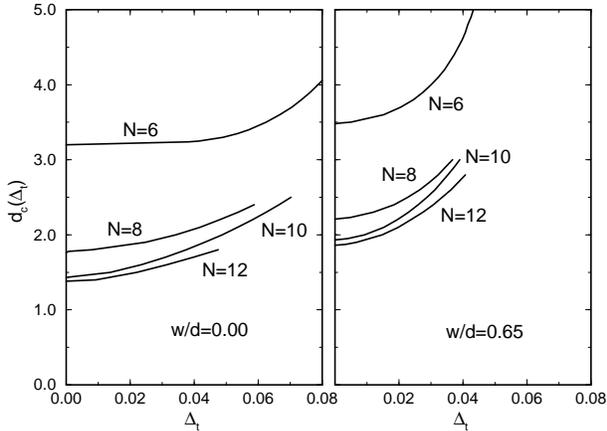}}
\caption{Phase boundaries for different system sizes $N$ and ratios of
well width to layer separation.
\label{fig2}}
\end{center}
\end{figure}

Figure \ref{fig2} shows the resulting phase boundaries for different 
system sizes and both cases of well width. At small layer separation the
system is in the ordered phase and the fluctuation peak occurs exactly at 
$\Delta_{t}=0$. At a critical layer separation $d_{c}(\Delta_{t}=0,N)$ the
phase boundary moves out rapidly to finite values of $\Delta_{t}$ and
intersects the axis $\Delta_{t}=0$ with an almost horizontal tangent. This
is in qualitative agreement with earlier experimental \cite{MEBPW:94} and
theoretical \cite{MPB:90} estimates of the phase diagram.

\begin{figure}
\begin{center}
\centerline{\includegraphics[width=8cm]{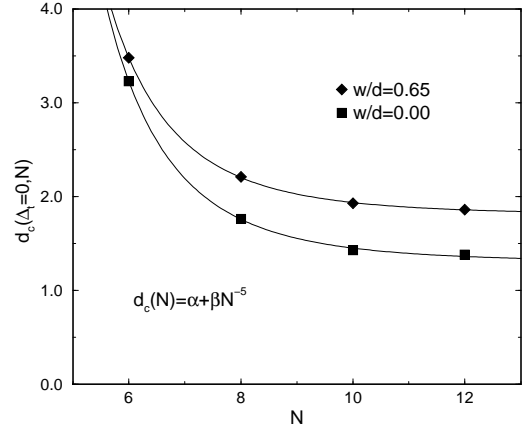}}
\caption{The critical layer separation $d_{c}(\Delta_{t}=0,N)$ (filled symbols)
at vanishing tunneling as a function of the system size $N$ for both cases of 
well width. The lines are finite-size fits to the data with
a shift exponent of $\lambda=5.0$.
\label{fig3}}
\end{center}
\end{figure}

The critical values $d_{c}(\Delta_{t}=0,N)$ form a rapidly
converging data sequence and are plotted in figure \ref{fig3}. 
These finite-size data are accurately and consistently 
described by an {\it ansatz} of the form 
$d_{c}(N)=\alpha+\beta N^{-\lambda}$ with two fit parameters 
$\alpha=d_{c}(N=\infty)$, $\beta$, and a shift exponent $\lambda$.
The best fits to both sets of data are obtained for 
$\lambda=5.0\pm 0.2$ leading to a value of $d_{c}(N=\infty)=1.30\pm 0.03$ for 
zero well width, and $d_{c}(N=\infty)=1.81\pm 0.03$ for $w/d=0.65$. 
The latter value is in excellent agreement with the results of Ref. 
\cite{SEPW:00}, where the onset of the tunneling conductance peak is observed
at a layer separation of $d=1.83$. Thus,
our numerical results clearly indicate that the findings of the above
tunneling experiments are the signature of a quantum phase transition.
The very large value of $\lambda$ seems inconsistent with a diverging 
correlation length and suggests the transition is first order.
A first order phase transition would explain the apparent 
coincidence of the appearance of spontaneous phase coherence and 
the quantum Hall effect in experiment \cite{MEBPW:94,SEPW:00}
We note that our result for the critical layer separation at vanishing 
tunneling gap agrees reasonably, at zero well width, with the point at which
the uniform density phase coherent state first becomes unstable in the 
Hartree-Fock approximation \cite{MPB:90}. 
At larger $w$, however, the Hartree-Fock estimates 
clearly deviate from the exact diagonalisation result.  

In order to further investigate the order of the quantum phase transition,
we 
introduce   
the ratio
\begin{equation}
\omega_{N}=\frac{2\left(\Delta T^{x}\right)^{2}_{N}}
{\left(d\langle T^{x}\rangle/d\Delta_{t}\right)_{N}}\,,
\label{omega}
\end{equation}
where the subscript $N$ refers to the system size. 
As we discuss below, this type of ratio should prove to be a powerful general 
tool in the analysis of any quantum phase transition.
In classical physics this ratio of fluctuation to susceptibility 
is equal to the thermal energy $k_B T$ and vanishes at $T =0$.  
The classical relationship does not apply here since  
the Hamiltonian fails to commute with its derivative with respect to
$\Delta_{t}$.   There is, however, a closely related zero-temperature
relationship with the typical excitation energy 
$\omega_{N}$ taking over the role of temperature.
The fluctuation can be written as
\begin{equation}
\left(\Delta T^{x}\right)^{2}=
\sum_{n>0}|\langle n|T^{x}|0\rangle|^{2}\,,
\label{fluc}
\end{equation}
where the sum is performed over all excited states, while for the derivative
of the pseudospin magnetisation one finds from linear response theory
\begin{equation}
\frac{d\langle T^{x}\rangle}{d\Delta_{t}}
=2\sum_{n>0}
\frac{|\langle n|T^{x}|0\rangle|^{2}}{E_{n}-E_{0}}\,.
\label{susc}
\end{equation}
From these equations we see that $\omega_N$ is a 
harmonic average of excitation energies $(E_{n}-E_{0})$, weighted by  
the factors $|\langle n|T^{x}|0\rangle|^{2}$. 
In particular, $\omega_{N}$
has a vanishing thermodynamic limit if at least one 
state with a nonvanishing matrix element
$\langle n|T^{x}|0\rangle$ has an
excitation energy $(E_{n}-E_{0})$ which 
extrapolates to zero for $N\to\infty$.  
Thus, equation (\ref{omega}) defines a characteristic energy scale of the
system at the phase boundary. The operator $T^{x}$ naturally enters
this expression since it couples to a control parameter driving the phase
transition. 

For a continuous phase transition one would clearly expect $\omega_{N}$ to
vanish at the phase boundary for an infinite system, while a finite limit 
$\lim_{N\to\infty}\omega_{N}$ is indicative of a finite energy scale, 
i.e. a first order transition. From our finite-size data for $\omega_{N}$
(evaluated at vanishing tunneling and $d=d_{c}(N)$) we conclude
that this quantity extrapolates for $N\to\infty$ 
to a rather substantial non-zero value 
of order $0.05 e^{2}/\epsilon l_{B}\sim 5$K for
both values of $w$ considered here. Along with the arguments and experimental
findings given so far, this result strongly suggests that the
bilayer quantum Hall system at filling factor $\nu=1$ undergoes a single
first order phase transition as a function of the ratio of layer separation
and magnetic length at all values of the tunneling 
amplitude. The phase boundary separates a phase with strong 
interlayer correlation (and a finite gap for charged excitations) from a 
phase with weak interlayer correlations and vanishing $E_{g}$.

Finally we comment briefly on the influence of a bias voltage between
the layers. 
When applying a bias voltage to the system the vector $\langle\vec T\rangle$
is tilted out of the $xy$-plane with a finite $z$-component. In this case
we find numerically that
the quantum phase transition is again signaled by the longitudinal 
fluctuation of the pseudospin magnetisation and its susceptibility,
and all results concerning the phase boundary are qualitatively the same. 
First order phase transitions from stronly correlated to weakly
correlated states also occur with increasing bias potential.  
We predict measureable anomalies in the double-layer system
capacitance at bias tuned phase transitions.  

We acknowledge helpful discussions with S. Sachdev. This work was 
supported by the Deutsche Forschungsgemeinschaft under Grant No. 
SCHL 539/1-1 and by the National Science Foundation under Grants No.
DMR-9714055 and DMR-0087133.


\begin{references}
\bibitem{note1} To compute the pseudopotentials of the Coulomb interaction
in the case of finite well width we take the electron wave function for
the growth direction in each well to be a normalised sine having its zeros
at the edges of the well. This means we neglect electron charge density
outside the wells, which is a very good approximation for not too large
tunneling $\Delta_{t}\lesssim 0.2(e^{2}/\epsilon l_{B})$, where the Coulomb 
energy scale is of order 100K . An illustrative picture of the charge
density distribution can be found in Ref. \cite{HDX:93}.
\end{references}
\end{document}